\documentclass[final,numberedheadings]{aipproc}

\usepackage{ }

\layoutstyle{6x9}

\newcommand{\hide}[1]{}
\renewcommand{\texttt}[1]{{\tt\small{#1}}}

\begin{document}

\title{A How-To for the Mock LISA Data Challenges}


\author{The \emph{Mock LISA Data Challenge Task Force}: Keith A. Arnaud}{address={Gravitational Astrophysics Laboratory, NASA Goddard Space Flight Center, \\ 8800 Greenbelt Rd., Greenbelt, MD 20771, USA}}
\author{Stanislav Babak}{address={Max-Planck-Institut f\"ur Gravitationsphysik (Albert-Einstein-Institut), \\ Am M\"uhlenberg 1, D-14476 Golm bei Potsdam, Germany}}
\author{John G. Baker}{address={Gravitational Astrophysics Laboratory, NASA Goddard Space Flight Center, \\ 8800 Greenbelt Rd., Greenbelt, MD 20771, USA}}
\author{Matthew J. Benacquista}{address={Center for Gravitational Wave Astronomy, University of Texas at Brownsville, \\ Brownsville, TX 78520, USA}}
\author{Neil J. Cornish}{address={Department of Physics, Montana State University, Bozeman, MT 59717, USA}}
\author{Curt Cutler}{address={Jet Propulsion Laboratory, California Institute of Technology, Pasadena, CA 91109, USA}}
\author{Shane L. Larson}{address={Center for Gravitational Wave Physics, 104 Davey Laboratory, University Park, PA 16802, USA},altaddress={Department of Physics, Weber State University, 2508 University Circle, Ogden, UT 84408, USA}}
\author{B. S. Sathyaprakash}{address={School of Physics and Astronomy, Cardiff University, Cardiff, CF243YB, UK}}
\author{Michele Vallisneri}{address={Jet Propulsion Laboratory, California Institute of Technology, Pasadena, CA 91109, USA}}
\author{Alberto Vecchio}{address={School of Physics and Astronomy, University of Birmingham, \\ Edgbaston, Birmingham B152TT, UK}}
\author{Jean--Yves Vinet}{address={Department ARTEMIS, Observatoire de la C\^ote d'Azur, BP 429, 06304 Nice, France}}



\begin{abstract}
The LISA International Science Team Working Group on Data Analysis (LIST-
WG1B) is sponsoring several rounds of mock data challenges, with the purpose
of fostering development of LISA data-analysis capabilities, and of
demonstrating technical readiness for the maximum science exploitation of the LISA data. The first round of challenge data sets were released at this
Symposium. We describe the models and conventions (for LISA and for gravitational-wave sources) used to prepare the data sets, the file format used to encode them, and the tools and resources available to support challenge participants.
\end{abstract}

\maketitle

The objectives, structure, and timeline of the Mock LISA Data Challenges (MLDCs) are discussed in the other contribution in this volume by the MLDC Task Force. Here we concentrate on the technical side of the challenges, and in particular on the theoretical models used to embody LISA and gravitational-wave (GW) sources in the first Challenge, and on the file format used to distribute the Challenge datasets.  More details can be found on the official MLDC website \cite{mldcweb}, in the living \emph{Omnibus} document for Challenge 1 \cite{omnibus}, and on the MLDC Task Force wiki \cite{mldcwiki}.

\section{Modeling LISA}

The analysis of real LISA data will necessarily involve a detailed modeling of instrument response and noise; for the purposes of the MLDCs, it is desirable to decouple this aspect from the inherent complexity of GW analysis, by distilling the LISA measurements into a standard idealized model.
Thus, the MLDC task force has developed a set of \emph{pseudo-LISA} assumptions and conventions, which we lay out in this section (see also Ref.\ \cite{omnibus}). 
Both of the LISA response simulators used in Challenge 1 (the LISA Simulator \cite{lisasimulator} and Synthetic LISA \cite{synthlisa}) comply with these assumptions, and adhere to these conventions. In later challenges, as the craft of LISA data-analysis matures, so will the pseudo-LISA model, becoming increasingly realistic.

\vspace{-0.4cm}
\subsection{The pseudo-LISA orbits}
\label{sec:orbits}

The pseudo-LISA orbits are obtained by truncating exact Keplerian orbits for a small mass orbiting the Sun to first order in the eccentricity (see the Appendix of Ref.~\cite{lisasimulator}). In Solar-System Barycentric (SSB) coordinates (with the $x$ axis aligned with the vernal point), we set
\begin{eqnarray}
x_n &=& a\cos \alpha + a \, e\left(\sin\alpha\cos\alpha\sin\beta_n
-(1+\sin^2\alpha)\cos\beta_n\right), \nonumber \\
y_n &=& a\sin \alpha + a \, e\left(\sin\alpha\cos\alpha\cos\beta_n
-(1+\cos^2\alpha)\sin\beta_n\right), \\
z_n & = & -\sqrt{3} \, a \, e \cos(\alpha-\beta_n) \, , \nonumber
\end{eqnarray}
where $\beta_n = (n-1)\times2\pi/3 + \lambda$ ($n=1, 2, 3$) is the relative orbital phase of each spacecraft, $a = 1$ AU is the semi-major axis of the guiding center, and $\alpha(t)=2 \pi \, t / (1 \mathrm{year}) + \kappa$ is its orbital phase. In this approximation, the spacecraft form a rigid equilateral triangle with side length $L = 2\sqrt{3} \, a \, e = 5\times 10^6$ km for $e=0.00965$. (In fact, the LISA Simulator and Synthetic LISA implement $e^2$-accurate orbits, but the additional terms make very little difference to the instrument response.)

The parameters $\kappa$ and $\lambda$ (\texttt{InitialPosition} and \texttt{InitialRotation} in lisaXML, see Sec.\ \ref{sec:lisaxml}) set the initial location and orientation of the LISA constellation; in Challenge 1, $\kappa=\lambda=0$. This choice places LISA at the vernal point, with spacecraft 1 directly below the guiding center in the southern ecliptic hemisphere. See Ref.\ \cite{omnibus} for expressions to convert to other LISA orbit specifications.

All times are measured by an ideal clock at the SSB.

\vspace{-0.4cm}
\subsection{The LISA response}
\label{sec:tdi}

The basic (individual-link) LISA response to GWs is taken to be the \emph{phase response} $\Phi_{ij}$ used in the LISA Simulator and discussed in Sec.\ II of Ref.\ \cite{lisasimulator}, or the \emph{fractional frequency response} $y^\mathrm{gw}_{slr}$ used in Synthetic LISA and discussed in Sec.\ II B of Ref.\ \cite{synthlisa}. (See the TDI Rosetta Stone \cite{rosetta} for translations between index notations.) The phase and fractional-frequency formalisms are equivalent, and are related by a simple time integration. The former has the advantage of representing more closely the actual output of the LISA phasemeters; the latter of being directly proportional to (differences of) the gravitational strains at the spacecraft.
(In fact the \emph{LISA Simulator} produces \emph{equivalent-strain} data, with a nominal length of $L_n = 10^{10}$ m. To convert equivalent strain to fractional frequency, differentiate and multiply by $2 \pi L_n / c$.)

LISA will employ Time-Delay Interferometry (TDI; see Refs.\ \cite{firstgen,modified,secondgen}) to cancel the otherwise overwhelming laser phase noise. In essence, TDI observables are constructed from time-delayed linear combinations of individual-link measurements, and they represent \emph{synthesized} interferometers where laser phase fluctuations move in closed paths across the LISA arms. More complicated paths are required to deal with the variations of the armlengths due to the finer details of the LISA orbits, giving rise to the three TDI ``generations.''

It is expected that high-level LISA data-analysis tasks (such as those targeted in the challenges) will be performed directly on TDI observables, and not on the underlying phase measurements. Thus, for the initial challenges we elect to represent the LISA output as \emph{TDI 1.5} observables \cite{modified,secondgen}, and in particular as the \emph{unequal-arm Michelson} observables $X$, $Y$, and $Z$ defined in Refs.\ \cite{secondgen}. Strictly speaking, TDI 2.0 would be required to cancel laser noise completely in a rotating and flexing LISA array such as our pseudo-LISA; however, the upgrade from TDI 1.5 to 2.0 changes little in the response to GW signals, but it requires more careful numerical treatments and adds to the complexity of analysis codes. Thus, the initial Challenge data sets contain TDI 1.5 observables \emph{without} laser noise.

\vspace{-0.4cm}
\subsection{The pseudo-LISA noises}
\label{sec:noises}

The model of LISA instrument noise adopted in Challenge 1 includes only contributions from optical noise (assumed white in phase, with one-sided spectral density 
$S_\mathrm{opt}^{1/2}(f) = 20 \times 10^{-12} \, \mathrm{m}\, \mathrm{Hz}^{-1/2}$), and from acceleration noise (assumed white in acceleration, but increasing as $1/f$ below $10^{-4}$ Hz, with one-sided spectral density $S_\mathrm{acc}^{1/2}(f) = 3 \times 10^{-15} [1 + (10^{-4}\,{\rm Hz}/f)^2]^{1/2}\, \mathrm{m}\, \mathrm{s}^{-2}\, \mathrm{Hz}^{-1/2}$), but not from laser phase noise, as discussed above.

The six optical noises and six acceleration noises (for the two optical benches on each spacecraft) are modeled as independent Gaussian random processes, and are realized in practice with sequences of pseudo-random numbers. Specifically, Synthetic LISA generates independent Gaussian deviates (i.e., white noise) in the time domain, and then filters them digitally to obtain the desired spectral shape; the LISA Simulator generates independent Gaussian deviates in the frequency domain, multiplies them by $S^{1/2}(f)$, and FFTs to the time domain.

\section{Modeling GW sources}
\label{sec:sources}

Another source of complexity that we wish to exclude from the initial challenges is the uncertainty about the true shape of the gravitational waveforms that Nature will provide to LISA. However, we can already begin prepare for their detection and analysis, while waiting for theory to provide more and more accurate models, by working with fully known waveforms of comparable structure and increasing complexity. This section describes the standard simplified waveforms used in Challenge 1 to embody the signals emitted by the three kinds of GW sources under consideration: galactic binaries, massive black-hole binaries, and extreme--mass-ratio inspirals. A special care was devoted to choosing standard source parametrizations that could be used by MLDC participants to report their analysis results and compare them easily. 

\vspace{-0.4cm}
\subsection{Conventions}
\begin{table}
\begin{tabular}{llll}
\hline
\tablehead{1}{l}{t}{Parameter} &
\tablehead{1}{l}{t}{Symbol} &
\tablehead{1}{l}{t}{Standard parameter name \\ (lisaXML descr.)} &
\tablehead{1}{l}{t}{Standard unit \\ (lisaXML descr.)} \\
\hline
Ecliptic latitude   & $\beta$   & \texttt{EclipticLatitude}  & \texttt{Radian} \\
Ecliptic longitude  & $\lambda$ & \texttt{EclipticLongitude} & \texttt{Radian} \\
Polarization angle  & $\psi$    & \texttt{Polarization}      & \texttt{Radian} \\
Inclination         & $\iota$   & \texttt{Inclination}       & \texttt{Radian} \\
Luminosity distance & $D$       & \texttt{Distance}          & \texttt{Parsec} \\
\hline
\end{tabular}
\caption{Common source parameters. Note that in the initial challenges we do not deal explicitly with the redshifting of sources at cosmological distances; thus, $D$ is a \emph{luminosity} distance, and the masses and frequencies of Tables \ref{tab:bbh} are those measured at the SSB, which are red/blue-shifted by factors $(1+z)^{\pm 1}$ w.r.t. to those measured locally near the sources.\label{tab:common}}
\end{table}

The sky location of a GW source is described by its J2000 \emph{ecliptic latitude} $\beta$ and \emph{longitude} $\lambda$, the latter measured from the vernal point, aligned with the $\hat{x}$ axis in our convention. We model gravitational radiation from the source as a plane wave traveling along the direction $\hat{k} = -(\cos \beta \cos \lambda, \cos \beta \sin \lambda, \sin \beta)$, with surfaces of constant phase given by $\xi = t - \hat{k} \cdot x$.
As written in the transverse--traceless gauge, the gravitational strain tensor can be decomposed in two standard polarization states,
\begin{equation}
\label{eq:defpol}
\mathbf{h}(\xi) = h_{+}(\xi) \left[ \hat{u}\otimes \hat{u} - \hat{v}\otimes \hat{v} \right] + h_{\times}(\xi) \left[ \hat{u}\otimes \hat{v} + \hat{v}\otimes \hat{u} \right],
\end{equation}
where $h_{+}(\xi)$ and $h_{\times}(\xi)$ multiply the polarization tensors ${\bf e}^{+}$ and ${\bf e}^{\times}$ formed from $\hat{u} = \partial \hat{k} / \partial{\beta}$, $\hat{v} \propto \partial \hat{k} / \partial{\lambda}$. Thus, GWs from any MLDC source are completely specified by $\beta$, $\lambda$, and by the two functions $h_+(\xi)$ and $h_\times(\xi)$ for the source's GW polarization amplitudes, measured at the SSB.

The orbital orientation of nonprecessing binaries is described by the inclination $\iota$ (the angle between the line of sight $-\hat{k}$ and the orbital angular momentum of the binary), and by their polarization angle $\psi$: specifically, if $h^S_{+}(\xi)$ and $h^S_\times(\xi)$ are the binary's GW polarizations in the source frame (i.e., defined with respect to the binary's \emph{principal polarization axes} $\hat{p}$ and $\hat{q}$) then 
\begin{equation}
h_+(\xi) + i h_\times(\xi) = e^{-2 i \psi} \left[ h^S_+(\xi) + i h^S_\times(\xi) \right],
\end{equation}
with $\psi = -\arctan(\hat{v} \cdot \hat{p} / \hat{u} \cdot \hat{p})$.
Together with $\beta$, $\lambda$, and with the luminosity distance $D$, $\iota$ and $\psi$ form a set of common standard parameters, listed in Tab.\ \ref{tab:common} with their standard lisaXML (see Sec.\ \ref{sec:lisaxml}) descriptors and units.

\vspace{-0.4cm}
\subsection{Galactic binaries}
Challenge 1 includes only searches for individually resolvable galactic binaries, as opposed to quasi-stochastic signals from populations of unresolvable sources. As an added simplification, all binaries are taken to be circular and monochromatic. Consequently, a Challenge-1 \texttt{GalacticBinary} source is completely determined by the parameters of Tables \ref{tab:common} and \ref{tab:galactic} together. The source-frame polarization amplitudes are computed in the restricted post-Newtonian approximation, and they are given by
\begin{eqnarray}
h^S_+(\xi)  & = & \mathcal{A} \left(1 + \cos^2{\iota}\right) \cos(2\pi f \xi + \phi_0), \\
h^S_\times(\xi) & = & -2 \mathcal{A} (\cos{\iota}) \sin(2\pi f \xi + \phi_0). \nonumber
\end{eqnarray}
The amplitude is specified explicitly among the source parameters; it is given in terms of the underlying physical parameters by $\mathcal{A} = (2 \mu / D) (\pi M f)^{2/3}$, with $M = m_1 + m_2$ the total mass, and $\mu = m_1 + m_2$ the reduced mass.
\begin{table}
\begin{tabular}{llll}
\hline
\tablehead{1}{l}{t}{Parameter} &
\tablehead{1}{l}{t}{Symbol} &
\tablehead{1}{l}{t}{Standard parameter name \\ (lisaXML descr.)} &
\tablehead{1}{l}{t}{Standard unit \\ (lisaXML descr.)} \\
\hline
Amplitude           & $\mathcal{A}$ & \texttt{Amplitude}    & \texttt{1} (GW strain) \\
Frequency           & $f$           & \texttt{Frequency}    & \texttt{Hertz} \\
Initial GW phase    & $\phi_0$      & \texttt{InitialPhase} & \texttt{Radian} \\
\hline
\end{tabular}
\caption{\texttt{GalacticBinary} source parameters. Note that \texttt{Amplitude} effectively replaces the standard \texttt{Distance} parameter.\label{tab:galactic}}
\end{table}

\vspace{-0.4cm}
\subsection{Massive--black-hole binaries}

For the sake of simplicity, all the massive--black-hole binaries GW sources considered in Challenge 1 are taken to be circular; black-hole spins are ignored, as are the final plunge and merger phases.
In such spin-less, circular, adiabatic binary inspirals, the Taylor-expanded post-Newtonian equations for energy balance can be integrated analytically, yielding expressions \cite{blanchet,DIS} for the orbital phase $\Phi$ as a function of the instantaneous orbital frequency $\omega$, and for the time to coalescence $t_c - t$ as a function of $\omega$. Truncating the two expressions to 2PN order, and inverting the second and substituting (numerically) in the first, we write the restricted post-Newtonian waveform for the inspiral as 
\begin{eqnarray}
h^S_{+}(\xi) &=& \frac{2\mu}{D}[M\omega(\xi)]^{2/3}(1+\cos^2 \iota)\cos [2\Phi(\xi)], \\
h^S_{\times}(\xi) &=& -\frac{2\mu}{D}[M\omega(\xi)]^{2/3}(2 \cos \iota) \sin [2\Phi(\xi)]. \nonumber
\end{eqnarray}
We end the waveform when one of the following conditions is realized: i) the (Schwarzschild) last stable orbit is reached \emph{or} ii) the ``MECO'' condition \cite{BCV2} is fulfilled \emph{or} iii) $\dot{\omega}$ becomes negative. Such a termination engenders ringing in the Fourier domain. In reality this would not happen, because the inspiral waveform flows smoothly into the plunge and merger waveforms (which we do not model). Thus, we smooth out the waveform, beginning at an orbital separation $R_\mathrm{taper} \in [7,9]M$, by multiplying it by the \emph{ad hoc} taper
\begin{equation}
w(t) = \left( 1 + \tanh\left[A (M/R - M/R_\mathrm{taper})\right] \right) / 2,
\end{equation}
where $R$ is approximated with Kepler's law ($R = M^{1/3} \omega^{-2/3}$), and
where the dimensionless coefficient $A = 150$ was determined empirically to produce smooth damping.

The lisaXML standard parameters of these Challenge-1 \texttt{BlackHoleBinary} sources are listed in Tables \ref{tab:common} and \ref{tab:bbh}.
\begin{table}
\begin{tabular}{llll}
\hline
\tablehead{1}{l}{t}{Parameter} &
\tablehead{1}{l}{t}{Symbol} &
\tablehead{1}{l}{t}{Standard parameter name \\ (lisaXML descr.)} &
\tablehead{1}{l}{t}{Standard unit \\ (lisaXML descr.)} \\
\hline
Mass of first BH    & $m_1$  & \texttt{Mass1}           & \texttt{SolarMass} \\
Mass of second BH   & $m_2$  & \texttt{Mass2}           & \texttt{SolarMass} \\
Time of coalescence & $t_c$  & \texttt{CoalescenceTime} & \texttt{Second} \\
Angular orbital phase & $\Phi_0$ & \texttt{InitialAngularOrbitalPhase} & \texttt{Radian} \\
\multicolumn{1}{r}{at time $t = 0$} & & & \\
Tapering radius & $R$    & \texttt{TaperApplied}    & \texttt{TotalMass} \\ 
\hline
\end{tabular}
\caption{\texttt{BlackHoleBinary} source parameters.\label{tab:bbh}}
\end{table}

\vspace{-0.4cm}
\subsection{Extreme--Mass-Ratio Inspirals}

The Extreme--Mass-Ratio Inspiral (EMRI) waveforms adopted in Challenge 1 are the Barack--Cutler ``analytic kludge'' waveforms \cite{BC}, whereby orbits are instantaneously approximated as Newtonian ellipses (and gravitational radiation is given by the corresponding Peters--Matthews formula \cite{pm}), but perihelion direction, orbital plane, semi-major axis, and eccentricity evolve according to post-Newtonian equations. While these waveforms are not particularly accurate in the highly relativistic regime of interest for real EMRI searches, they do exhibit the main qualitative features of the true waveforms, and they are considerably simpler to generate.  It is expected that any search strategy that works for them could be modified fairly easily to deal with the true general-relativistic waveforms, once these become available.

The ``analytic kludge'' waveforms are too complex to describe in this restricted space, so we refer the reader to Refs.\ \cite{BC} and \cite{omnibus}, and we content ourselves with presenting a complete table of Challenge-1 \texttt{EMRI} parameters in Tab.\ \ref{tab:emri}.
\begin{table}
\begin{tabular}{llll}
\hline
\tablehead{1}{l}{t}{Parameter} &
\tablehead{1}{l}{t}{Symbol} &
\tablehead{1}{l}{t}{Standard parameter name \\ (lisaXML descr.)} &
\tablehead{1}{l}{t}{Standard unit \\ (lisaXML descr.)} \\
\hline
Mass of central BH     & $M$      & \texttt{MassOfSMBH}           & \texttt{SolarMass} \\
Mass of compact object & $\mu$  & \texttt{MassOfCompactObject}  & \texttt{SolarMass} \\
Central-BH spin        & $|S|/M^2$  & \texttt{CoalescenceTime}      & \texttt{Second} \\
Central-BH spin orientation & $\theta_K, \phi_K$ & \texttt{PolarAngleOfSpin}, & \texttt{Radian} \\
 \multicolumn{1}{r}{w.r.t. SSB frame}  &                         & \texttt{AzimuthalAngleOfSpin} & \\ \hline
& & \texttt{InitialAzimuthal}\ldots & \\                   
Azimuthal orbital freq.\ at $t = 0$ & $\nu_0$ & \multicolumn{1}{r}{\ldots\texttt{OrbitalFrequency}} & \texttt{Hertz} \\
Azimuthal orbital phase at $t = 0$ & $\Phi_0$ & \multicolumn{1}{r}{\ldots\texttt{OrbitalPhase}} & \texttt{Radian} \\ \hline
Eccentricity at $t = 0$ & $e_0$ & \texttt{InitialEccentricity} & \texttt{1} \\
Direction of pericenter at $t = 0$ & $\tilde{\gamma}_0$ & \texttt{InitialTildeGamma} & \texttt{Radian} \\
Direction of orbital angular & $\alpha_0$ & \texttt{InitialAlphaAngle} & \texttt{Radian} \\
\multicolumn{1}{r}{momentum w.r.t. $S$ at $t = 0$} & & & \\
Opening angle & $\lambda$ & \texttt{LambdaAngle} & \texttt{Radian} \\ 
\hline
\end{tabular}
\caption{\texttt{EMRI} source parameters. Note that EMRIs do not use the nonprecessing-binary inclination $\iota$, and be aware of the collision between the symbols for the EMRI compact-object mass ($\mu$) and opening angle ($\lambda$), the binary reduced mass (again $\mu$), and the ecliptic longitude (again $\lambda$).\label{tab:emri}}
\end{table}

\section{Encoding challenge data sets}

All MLDC training and challenge data sets are distributed from the MLDC website \cite{mldcweb} in a standard file format, developed by the MLDC Task Force with the goal of facilitating the use of the data sets on different computing platforms, and of enabling their identification, perusal, tracking, and archival. The MLDC file format is also used internally by the MLDC Task Force within the workflow that leads from the choice of random source parameters to the generation of TDI data sets.

\vspace{-0.4cm}
\subsection{File structure}

\begin{figure}
  \includegraphics[height=.3\textheight]{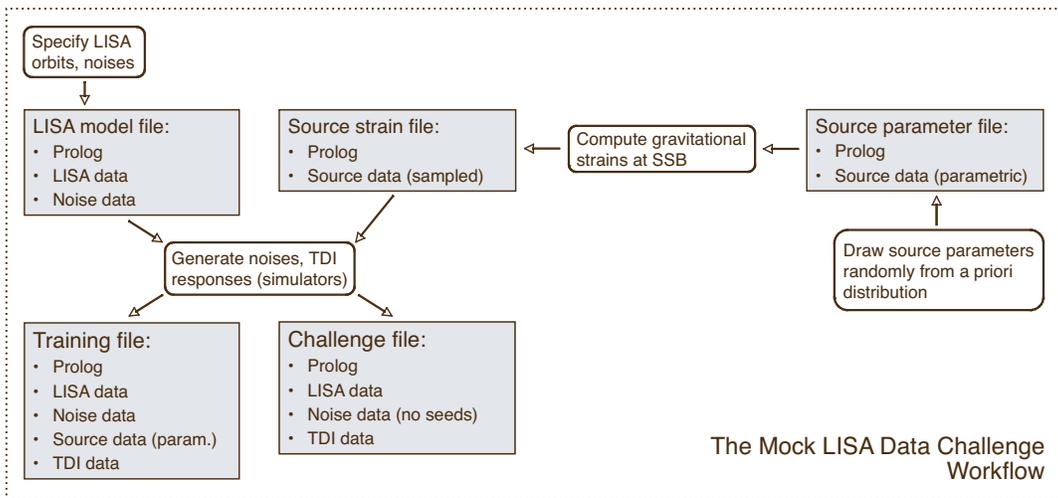}
  \caption{The MLDC workflow: the final data products (the training and challenge data sets) and the intermediate work data are all represented as MLDC files that include different combinations of the possible sections.\label{fig:workflow}}
\end{figure}

Depending on its use, a MLDC file will contain one or more of the following sections:
\begin{itemize}
\item A \emph{prolog} including file metadata such as author, generation date, the name and version of the computer code used to create the data, and any other relevant comments.
\item A \emph{LISA data} section describing the model of the LISA orbits used in the simulations; for the initial challenges, this amounts to the \texttt{InitialPosition} and \texttt{InitialRotation} needed to fully specify the pseudo-LISA orbits of Sec.\ \ref{sec:orbits}.
\item A \emph{noise data} section describing the models of the LISA noises used in the simulations; for the initial challenges, this amounts to the power spectral densities and generation timesteps of the pseudo-random sequences described in Sec.\ \ref{sec:noises}.
%
\item A \emph{source data} section describing the gravitational waveform(s) included in the simulation. The sources may be specified in terms of the standard source parameters defined in Sec.\ \ref{sec:sources}; otherwise, they may be represented as explicit time series of the $h_+$ and $h_\times$ gravitational strains at the SSB, plus a minimal set of parameters including the source's sky position and the time offset, cadence, and length of the time series.
\item A \emph{TDI data} section containing one or more time series of TDI observables, assembled from LISA's response to noises and/or sources, as described in Sec.\ \ref{sec:tdi}; for the initial challenges, the observables of choice are $X$, $Y$, and $Z$ of TDI 1.5, plus (trivially) the SSB time.
The standard names of these observables are \texttt{Xp}, \texttt{Yp}, and \texttt{Zp} for the equivalent-strain version of the data sets (generated with the LISA Simulator), and \texttt{Xf}, \texttt{Yf}, and \texttt{Zf} for the fractional-frequency--fluctuation version (generated with Synthetic LISA).
\end{itemize}
Thus, training data sets will be represented by MLDC files including all types of sections; on the other hand, challenge data sets (which must be ``blind'') will omit the source-data section.
Different combinations of the sections appear in the intermediate files used in the MLDC workflow, shown in Fig.\ \ref{fig:workflow}. 


\vspace{-0.4cm}
\subsection{Implementation}
\label{sec:lisaxml}

The MLDC file format is implemented using XML (the eXtensible Markup Language), a simple, flexible text format related to HTML, and widely used in the exchange, manipulation, and storage of many kinds of data, especially on the world-wide web \cite{xml}. Software libraries to handle XML are readily available for most computer languages.
XML documents consist of a nested hierarchy of \emph{elements}, enclosed by opening and closing \emph{tags} (\texttt{<tagname>} and \texttt{</tagname>}), and containing textual data; elements may also have textual \emph{attributes} (as in \texttt{<tagname attrname="attrvalue">}).
White space and newlines have no meaning within XML (any sequence of such characters is in fact equivalent to a single white space), but they are usually added liberally to help human parsing of the files.
%

The XML implementation of the MLDC file format (known as lisaXML) is based on XSIL (the eXtensible Scientific Interchange Language), an XML dialect developed at Caltech to represent scientific data in multiple applications \cite{xsil}. XSIL is very terse; its data structures consist of few simple building elements:
\begin{itemize}
\item \texttt{<XSIL Name="..." Type="...">} acts as a hierarchical container. For instance, the LISA-data section of an MLDC file is represented by an \texttt{<XSIL Type="LISAData">} element, and the source-data section is represented by a \texttt{<XSIL Type="SourceData">} element containing one or more \texttt{<XSIL Name="..." Type="PlaneWave">} elements.
\item \texttt{<Param Name="..." Unit="...">} is used to describe parameter values and their units. For GW sources, the \texttt{Name} and \texttt{Unit} attributes are those found in Tables \ref{tab:common}--\ref{tab:emri}. For instance, 
\texttt{GalacticBinary} may have \texttt{<Param Name="Frequency" Unit="Hertz">1.0e-3</Param>}.
\item \texttt{<Array Name="..." Type="...">} is used to specify arrays of homogeneous data, such as the time series of TDI observables. In lisaXML, the actual array elements are stored in separate binary files, but the \texttt{Array} element contains information about the location, organization, and encoding of the binary files.
\end{itemize}
The advantage of this hybrid XML/binary arrangement is that bulk data are represented in binary format, which can be saved, stored, and loaded very efficiently, while file metadata and the LISA, noise, and source parameters are contained in textual XML files that are easily parseable and editable by humans (or by powerful XML libraries). In fact, lisaXML files can be viewed in standard-compliant web browsers such as Firefox and Safari, which will use a set of special lisaXML \emph{stylesheet transformations} \cite{xsl} to render the metadata and nested parameters structures (but not the bulk data!) as pleasantly formatted tables. 

\vspace{-0.4cm}
\subsection{Usage}

The MLDC Task Force has developed dedicated software tools to read and write lisaXML files from different computing environments:
\begin{itemize}
\item \emph{C/C++.} The LISA Tools Subversion archive \cite{lisatools} includes a directory \texttt{lisaXML/io-C}, which contains lisaXML input--output routines in C; TDI and strain time series are returned as simple C arrays, while source and time series parameters are stored in C structures.
\item \emph{Python.} Routines to input and output lisaXML are built into the Python-steered Synthetic LISA \cite{synthlisa}, which can read (write) strain and TDI time series into (from) standard NumPy arrays \cite{numpy}. In addition, the Synthetic LISA Python/C++ objects that describe the LISA geometry, the LISA noises, and GW sources can also be translated to and from their lisaXML representation.
%
\item \emph{MATLAB.} The program \texttt{xml2matlab.c}, found in directory \texttt{lisaXML/C-exam}\-\texttt{ples} in the LISA Tools Subversion archive \cite{lisatools}, can be compiled into a MATLAB MEX function that will read lisaXML TDI time series into a MATLAB array.
\item \emph{ASCII.} The program \texttt{xml2ascii.c}, in the same location, can be compiled into a command-line utility that will dump lisaXML TDI time series into a tab-separated ASCII file, or to standard output.
\end{itemize}
All these tools are still in development at the time of writing, but all are already quite functional in reading MLDC data sets into analysis applications.
For more detailed instructions on reading lisaXML files, see the MLDC omnibus documents \cite{omnibus} and the LISA Tools SourceForge website \cite{lisatools}; these
include also the instructions and scripts necessary to reproduce the MLDC workflow and create additional training sets. 


\begin{theacknowledgments}
M.V.'s work was supported by the LISA Mission Science Office and by the Human Resources Development Fund at the Jet Propulsion Laboratory, California Institute of Technology, where it was performed under contract with the National Aeronautics and Space Administration.
\end{theacknowledgments}

\end{document}